\documentclass[aps,prl,twocolumn,groupedaddress,showpacs]{revtex4}

\usepackage{graphicx}

\begin{document}
\newcommand{\beq}{\begin{equation}}
\newcommand{\eeq}{\end{equation}}
\newcommand{\etal}{{\textit{et al.~}}}

\title{Critical dislocation speed in helium-4 crystals}

\author{Ariel Haziot, Andrew D. Fefferman, Fabien Souris, John R. Beamish*, Humphrey J. Maris**, and S\'ebastien Balibar}

\affiliation{Laboratoire de Physique Statistique de l'ENS, associ\'{e} au CNRS et aux
Universit\'{e}s Denis Diderot et P.M Curie, 24 rue Lhomond 75231 Paris Cedex
05, France\\
* Permanent address: Department of Physics, University of Alberta, Edmonton, Alberta, Canada T6G 2E1\\
** Permanent address: Department of Physics, Brown University, Providence, RI 02912, USA}

\date{\today}

\begin{abstract}
Our experiments show that, in $^4$He crystals,  the binding of $^3$He impurities to dislocations does not necessarily imply their pinning. Indeed, in these crystals, there are two different regimes in the motion of dislocations when impurities bind to them. At low driving strain $\epsilon$ and frequency $\omega$ , where the dislocation speed is less than a critical value (45$\mu$m/s), dislocations and impurities apparently move together. Impurities really pin the dislocations only at higher values of $\epsilon\omega$. The critical speed separating the two regimes is  two orders of magnitude smaller than the average speed of free $^3$He impurities in the bulk crystal lattice. We obtained this result by studying the dissipation of dislocation motion as a function of the frequency and amplitude of a driving strain applied to a crystal at low temperature.
Our results solve an apparent contradiction between some experiments, which showed a frequency-dependent transition temperature from a soft  to a stiff state, and other experiments or models where this temperature was assumed to be independent of frequency.  The impurity pinning mechanism for dislocations appears to be more complicated than previously assumed.
\end{abstract}


\maketitle

\begin{center}
\textbf{Introduction}
\end{center}

Extending earlier work by Paalanen \etal\cite{Paalanen81}, Day and Beamish \cite{Day07} studied an elastic anomaly in solid helium 4, which is now generally accepted  as  a ``giant plasticity'' due to the motion of dislocations. It implies a strong reduction in the crystal stiffness  around 200~mK, between binding of dislocations to $^3$He impurities at low temperature and damping of their motion by collisions with thermal phonons at higher temperature \cite{Day10,Syshchenko10,Rojas10,Varoquaux12,Haziot13a,Haziot13b}. The precise motion of the dislocations has been identified as a gliding parallel to the basal planes of the hexagonal structure of these hcp crystals \cite{Haziot13a}, so that the anomalous softening is highly anisotropic in single crystals. An analysis of the damping by thermal phonons has allowed a determination of the dislocation density $\Lambda$ and of the free length $L_{\mathrm N}$ between nodes in the dislocation network \cite{Haziot13b}. As a consequence,  a precise study of the elastic response to an applied strain allows the determination of the dislocation speed. Moreover, several authors \cite{Beamish12,Maris12,Reppy12-175,Chan12} have also realized that this elastic anomaly is the likely explanation for the rotation anomaly of crystals in ``torsional oscillators'', which was previously attributed to supersolidity \cite{Kim04a,Kim04b}.

As explained in this article, we have found evidence for a characteristic speed below which dislocations can move with $^3$He impurities attached to them. We call it ``critical speed'' for simplicity but it should be clear that there is no analogy with critical phenomena nor with the critical speed beyond which dissipation occurs in a superfluid flow. Our new results solve a running controversy about the frequency dependence of this motion, some authors \cite{Syshchenko10} having observed that the transition  temperature from a soft to a stiff state depends on frequency while some others have assumed \cite{Iwasa10,Kang13} that there is no frequency dependence. Furthermore, our new observations appear important to understand the binding mechanism of $^3$He impurities to dislocation and the damping of their motion in the low temperature regime where impurities bind to them.

As concerns how $^3$He binding affects the crystal stiffness, two different mechanisms have been considered. Iwasa \cite{Iwasa10} and Kang \etal \cite{Kang13} proposed that the transition from a soft to a stiff state is only a consequence of the free length $L$ of the dislocation evolving from a network pinning length $L_{\mathrm N}$ to a smaller length, an impurity pinning length $L_{\mathrm i}$. Within the Granato-Lucke  theory \cite{Granato56}, at low frequency and if the dislocation damping is small, the temperature dependent shear modulus $\mu$ is given by
\beq
\label{GL}
  \frac{\Delta\mu}{\mu_{\rm{el}}}= \frac{\mu_{\rm{el}} - \mu}{\mu_{\rm{el}}} = \frac{R \Sigma \Lambda L^{2}}{1+R \Sigma \Lambda L^{2}} \: ,
\eeq
where $\mu_{\rm{el}}$ is the  intrinsic value of the shear modulus corresponding to the stiff state when dislocations do not move; $R$ is the orientation factor ($0\leq R \leq 0.5$) which relates the shear stress in the dislocations' glide direction to the applied stress and can be determined if the crystal orientation is known;  $\Sigma =4 (1 - \nu)/\pi^3\approx 0.09$ ($\nu$ is Poisson's ratio).  In this approach, the transition temperature does not depend on frequency.

In another approach, Syshchenko \etal  proposed a Debye model where the crystal properties have a relaxation time $\tau$ that depends exponentially on temperature due to a binding energy $E$ of dislocations to $^3$He impurities. The transition was understood to be frequency dependent because it  takes place where  $\omega\tau$~=~1, $\omega$ being the angular frequency of the applied strain. From an Arrhenius plot of the peak dissipation temperature, Syshchenko \etal found $E$ respectively equal to 0.73 and 0.77~K in two different samples, more precisely a distribution of binding energies around these values. The energy distribution was found necessary to fit  the temperature variation of both the stiffness amplitude and the associated dissipation.

In order to make progress in the interpretation of the elastic anomaly of helium crystals, we have studied the shear modulus of one particular single crystal at variable frequency and strain amplitude. As described below, we have discovered a crossover from a frequency dependent regime to a frequency independent regime. It suggests that $^3$He impurities could follow the dislocation motion only at velocities less than a critical value $v_{\mathrm c} \approx 45~\mu$m/s. Note that the critical speed we observe is not  analogous to a critical speed in a superfluid flow. It is a change in the dissipation mechanism, not an onset of dissipation.

\begin{figure}
\includegraphics[width=\linewidth]{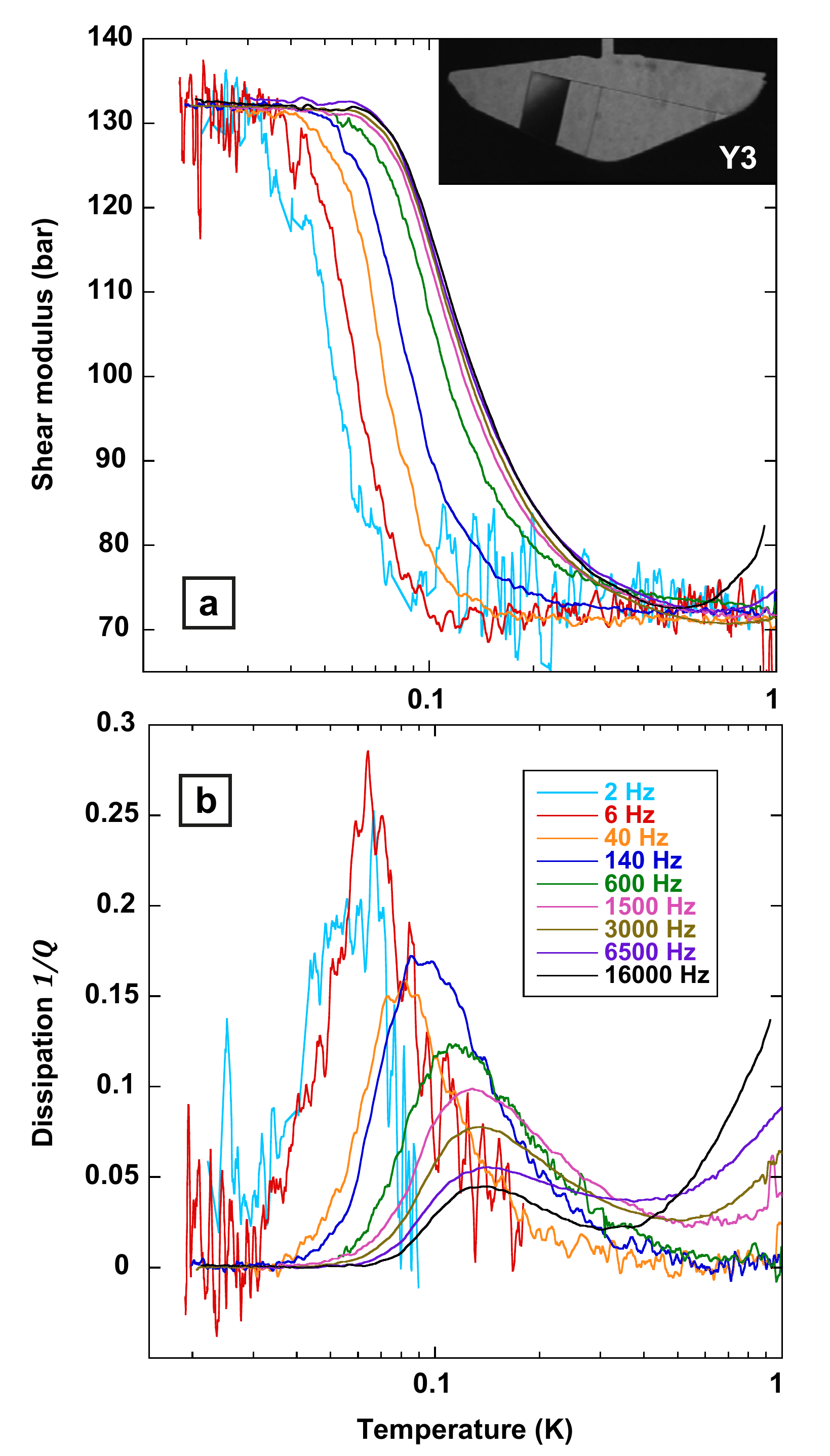}
\caption{(Color online) The shear modulus (a) and the dissipation (b) from the crystal response to a driving strain $\epsilon =  2.7~10^{-9}$.  The different curves correspond to different  frequencies. On the top right corner, a picture of the crystal Y3 shows a facetted shape with a c-axis tilted by 12.5$^{\circ}$ from the vertical.}
\label{fig:Figure1}
\end{figure}

\begin{center}
\textbf{Experimental setup}
\end{center}

Our experimental cell has already been described in previous articles \cite{Haziot13a,Haziot13b}. It is a 5~cm$^3$ hexagonal hole in a 15~mm thick copper plate, which is closed by two sapphire windows. Inside the cell,  two piezo-electric transducers face each other with a narrow gap in between. The gap thickness is $d$~=~0.7~mm. The cell is attached to the mixing chamber of a dilution refrigerator with an optical access and the crystal orientation is determined from its growth shape. An AC-voltage $V$ is applied to one transducer in order to produce a strain at a frequency $f = \omega/2\pi$ in the range 1~Hz to 20~kHz. The  displacement of the transducer surface is
$d_{15} = 0.95~\rm\AA/V$ so that applying a voltage in the range 1~mV to 1~V, produces a strain
 $\epsilon$ between 10$^{-10}$ and 10$^{-7}$.  The  stress $\sigma$ on the other transducer is  $\sigma = \mu\epsilon$  in the range 10$^{-9}$ to 10$^{-5}$~bar depending on the value of the relevant shear modulus $\mu$ (20 to 150~bar). By measuring the amplitude  and the phase $\Phi$  of the response to the applied strain, we obtain the shear modulus $\mu$ and the dissipation $1/Q = \tan{\Phi}$.

For the particular study that is presented here, we chose a  particular $^4$He single crystal named Y3, which was grown at 1.35 K from natural purity $^4$He containing 0.3~ppm of $^3$He. After growth, it was  cooled along the melting line down to low temperature in order to crystallize some remainder of liquid  in  cell corners \cite{Haziot13a} and to have a temperature independent $^3$He concentration equal to that in natural $^4$He. By analyzing the picture shown  on  Figure 1, we found that the c-axis was tilted by  $\theta = 12.5^{\circ}$ from the vertical and its projection in the horizontal plane by $\phi = 90^{\circ}$ from the normal to the cell windows.  With this orientation and a vertical shear, one measures a shear modulus  $\mu=0.82c_{44}+0.04(c_{11}+c_{33}-2c_{13})$ that mainly depends on the elastic coefficient $c_{44}$. This is the coefficient which varies with T when dislocations glide along the basal planes \cite{Haziot13a}. As explained in Ref.~\cite{Haziot13a}, the dislocations are pinned at low temperature so that all elastic coefficients including $c_{44}$ equal the values measured at high T and high frequency  by Crepeau \etal \cite{Crepeau} and by Greywall \cite{Greywall}. In the present case, it implies a stiff state at low T with a shear modulus $\mu_{\rm el}$ = 135~bar. More details on the experimental setup and the growth techniques we used are given in Ref.~\cite{Haziot13a}, especially its Supplementary Material.

\begin{center}
\textbf{Results}
\end{center}
\begin{figure}
\includegraphics[width=\linewidth]{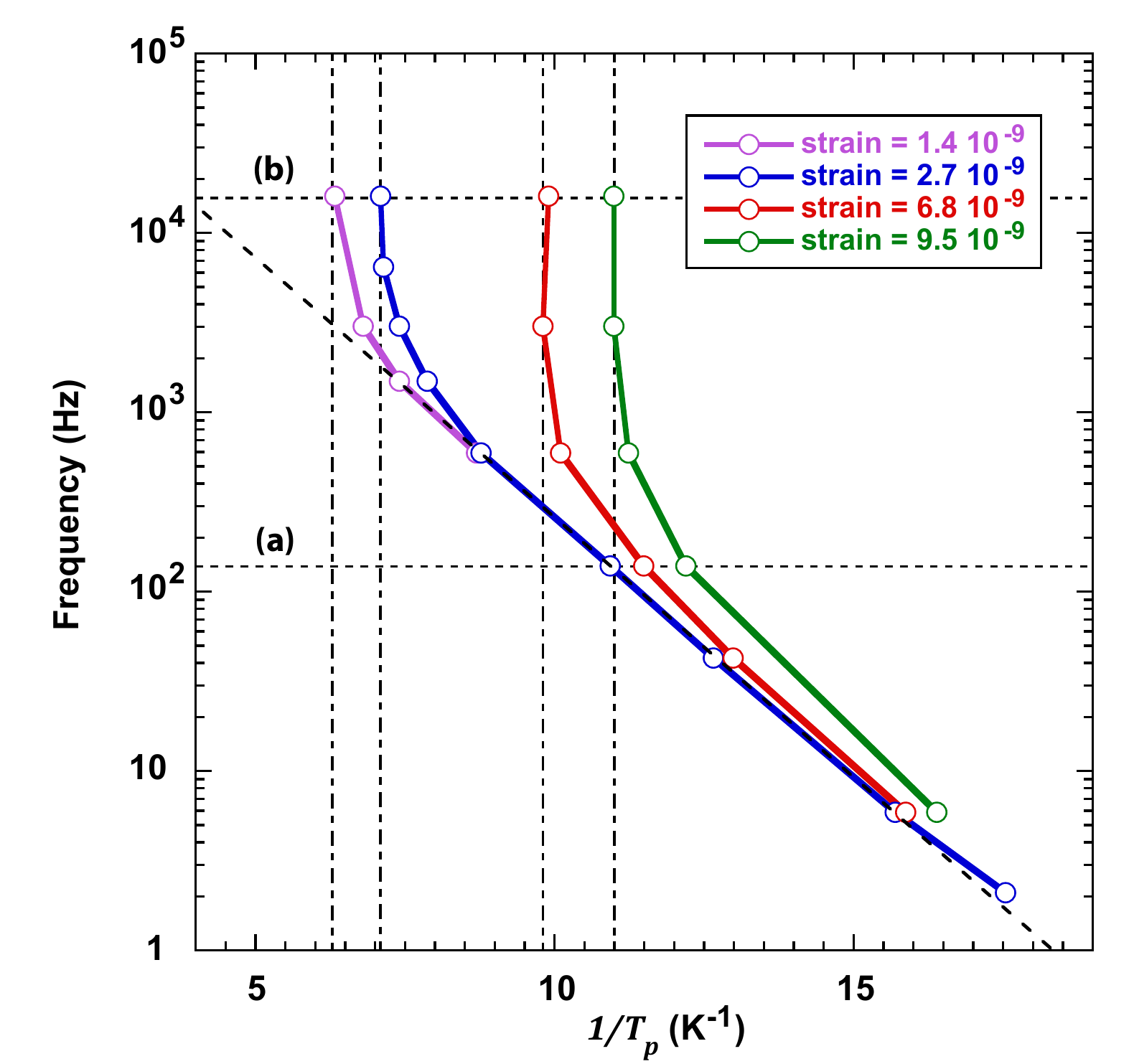}
\caption{(Color online) A semi-log Arrhenius plot of the measurement frequency as a function of the inverse of the temperature $T_{\mathrm p}$ at which the dissipation reaches its peak value. $T_{\mathrm p}$  represents the transition temperature from the soft to the stiff state of the crystal. Error bars on the determination  $T_{\rm p}$ are less than symbol sizes even at the lowest frequency and amplitude. The linear regime at low frequency  indicates the existence of a thermally activated process with an activation energy $E$=0.67 K.}
\label{fig:Figure2}
\end{figure}

\begin{figure}
\includegraphics[width=\linewidth]{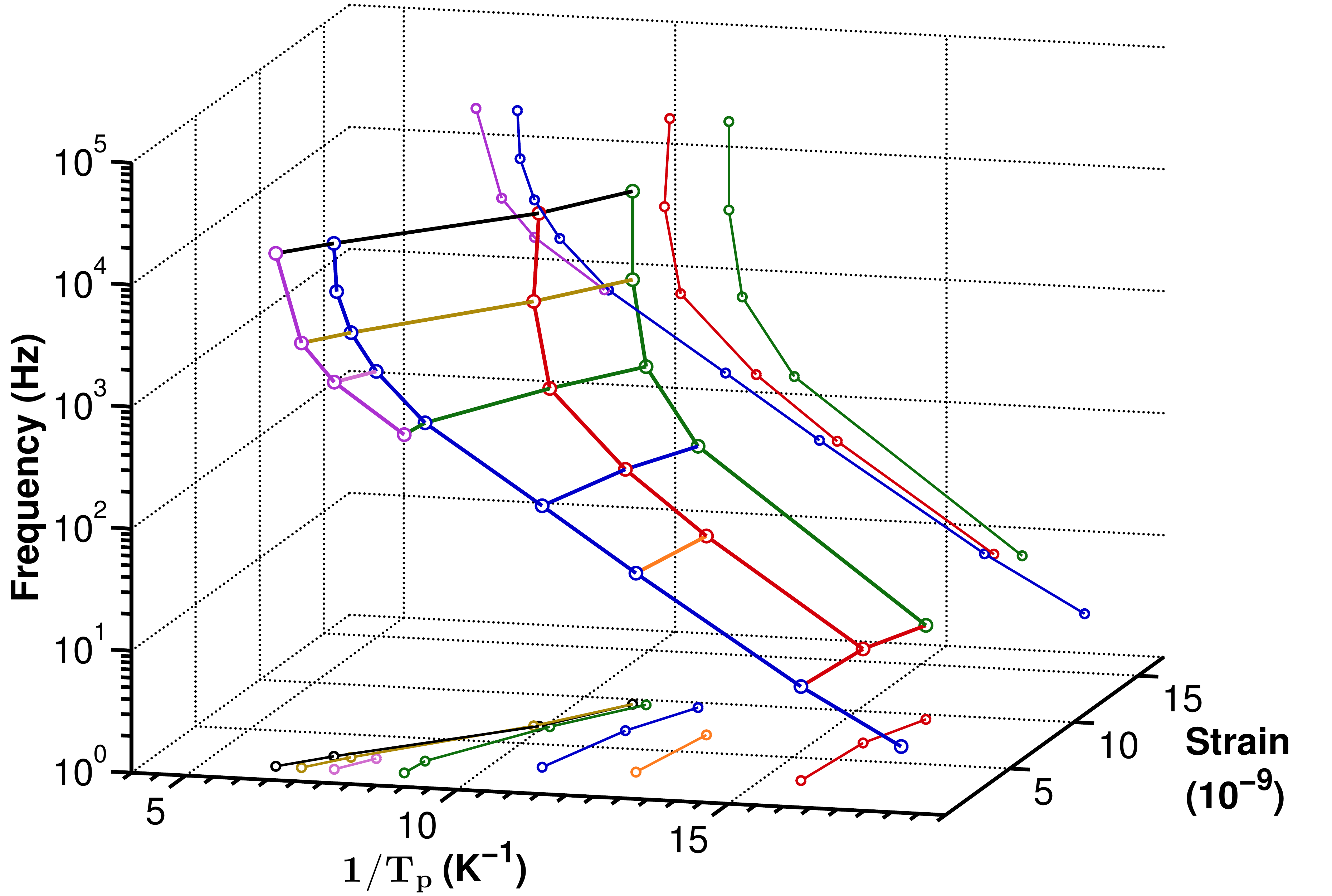}
\caption{(Color online) Three dimensional plot of the inverse transition temperature $1/T_{\mathrm p}$ as a function of both the frequency and the amplitude of the applied strain. A projection of the plot is shown in the frequency/temperature plane, which is shown more clearly on Fig.~2. Another projection is shown on this figure, now in the strain/temperature plane, which has been used to build Fig.~5.}
\label{fig:3DPlotStress}
\end{figure}
We have measured the shear modulus and the dissipation as a function of $T$ by cooling the crystal from 1~K down to 20~mK. We used  frequencies from 2 Hz to 16 kHz. We varied the driving strain from 1.4~10$^{-9}$ to 9.5~10$^{-9}$. We verified that this driving strain is sufficiently small for the crystal to reach its fully stiff state at 20 mK. Above 10$^{-8}$ it would no longer be true.

Figure~1 shows the shear modulus and the dissipation measured at a strain $\epsilon$ = 2.7 10$^{-9}$ for several frequencies. The transition from the stiff to the soft state occurs at a temperature that increases with frequency. This result confirms the previous observations by  Syshchenko \etal\cite{Syshchenko10} who worked at small driving strain. The low temperature stiffness is independent of frequency. The minimum value of the shear modulus is also frequency independent, corresponding to a  softening by 60\%, in good agreement with previous measurements on similar crystals \cite{Haziot13a}. What depends on frequency  is the transition temperature and the amplitude of the dissipation near this transition temperature.

Figure~2 shows the frequency dependence of the dissipation peak temperature, which can be considered as the transition temperature. The semi-log  ``Arrhenius plot'' shows a linear dependence in $1/T_{\mathrm p}$, which means the existence of a thermally activated regime at low frequency and low strain amplitude. We have found an activation energy E = 0.67~K, which we propose to identify with the binding energy of $^3$He impurities to dislocations, as we explain at the end of this article. A similar behavior was observed by Syshchenko \etal who supposed that the relaxation time $\tau$ of dislocations was proportional to the concentration $X_{\rm 3}^{\rm d} \propto \exp{(E/k_{\rm B} T)}$ of impurities bound to them. Assuming that $\omega = 1/ \tau $ at the peak dissipation, they found  $^3$He binding energies  $E$ respectively equal to 0.73 and  0.77~K in two different samples \cite{Syshchenko10}.

Before discussing this interpretation, let us remark that our measurements are more precise so that they show two additional properties:

1- the transition temperature is independent of strain amplitude in the low frequency and low amplitude limits. This indicates that the concentration of $^3$He bound to the line  depends on temperature only. But at higher frequency and amplitude the transition temperature  depends on strain amplitude, which now indicates that applying a large strain lowers the temperature where  $^3$He impurities bind to the dislocations. Fig.~3 shows the 3-dimensional plot of $1/T_{\mathrm p}$ as a function of frequency and applied strain. Fig.~2 is a projection of this 3D-plot.

2-  Above a certain frequency, which varies with the strain amplitude $\epsilon$, the transition temperature becomes independent of frequency. The crossover from a frequency independent regime to a frequency  dependent regime had not been observed before.

Figure~4 shows a graph similar to that of Figure~2 where the frequency $\omega$ has been replaced by the velocity of the dislocation line, which is calculated after assuming that its pinning length is the network pinning length $L_{\mathrm N}$. This new graph shows that the crossover between the two regimes occurs at a constant critical velocity $v_{\mathrm c} \approx 45\: \mu$m/s. In order to obtain the velocity $v$ we have used our previous work \cite{Haziot13b} as follows.

We first calculated the dislocation density $\Lambda$ and the free length $L$ in their network from eq.~1 and the following one, which relates the dissipation $1/Q$ to the same quantities $\Lambda$ and $L$:

\beq
  \frac{1}{Q} =\frac{\Delta\mu}{\mu_{\rm{el}}} \omega\tau = \frac{\Delta\mu}{\mu_{\rm{el}}} \omega \frac{B L^{2}}{\pi^{2} C} .
\label{eq:1/Q}
\eeq
where the dislocation energy per unit length is
\beq
C = \frac{2 \mu_{\rm{el}} b^{2}}{\pi (1-\nu)}
\eeq
$\mu_{\rm{el}}$ is the intrinsic elastic shear modulus in the absence of dislocation motion, $b = 0.37$~nm is the magnitude of the Burgers vector,  and $\nu = 1/3$ is Poisson's ratio. The coefficient $B$ is the damping of the dislocation motion, which is due to $^3$He atoms at low temperature but to collisions with thermal phonons at high temperature where it reads:
\beq
  B = \frac{14.4 {k_B}^3}{{\pi}^2 {\hbar^2}{c^3}} T^3
\label{eq:B}
\eeq
with $c$ the Debye velocity of sound.
The length $L$ equals $L_{\mathrm N}$ in the giant plasticity domain where the shear modulus reaches its minimum value.

\begin{figure}
\includegraphics[width=\linewidth]{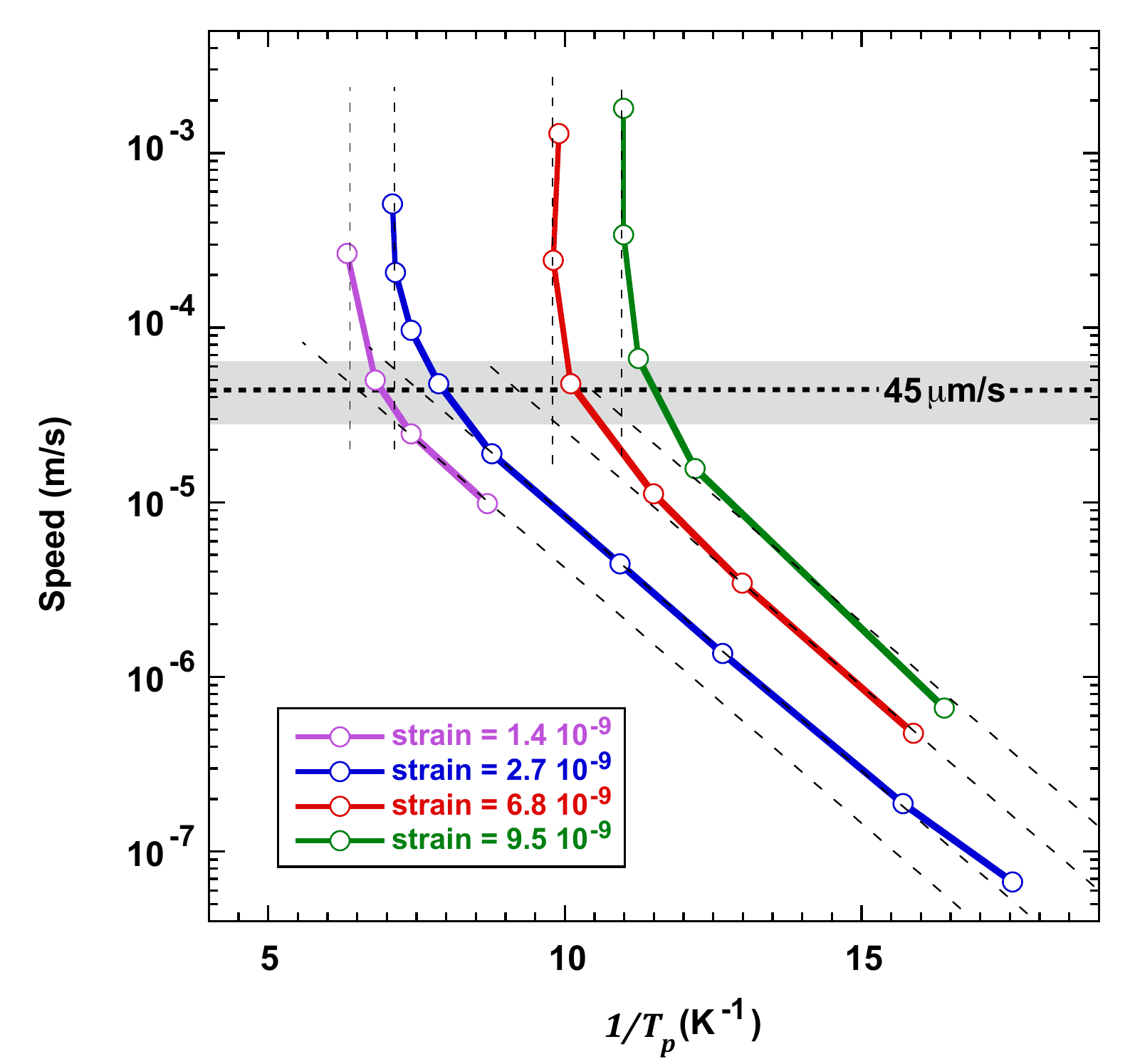}
\caption{(Color online) This graph shows the same data set as on Figure 2 except that the frequency has been replaced by the maximum speed of the dislocations (see text for its calculation). This representation shows that the crossover between the two regimes occurs at a constant critical speed $v_{\mathrm c} \approx 45 \mu$m/s.}
\label{fig:Figure4}
\end{figure}

From eqs.~\ref{GL} , \ref{eq:1/Q} and \ref{eq:B}, we found a network pinning length $L_{\mathrm N}=73\: \mu$m and a density $\Lambda=7.6\:\:10^{5}$cm$^{-2}$ for the single crystal Y3 studied in this paper. This gives a factor $\Lambda L_{\mathrm N}^2$ = 40, which means a small connectivity of the dislocation network and explains the large variation of the shear modulus \cite{Friedel64}. We then derived the maximum dislocation speed from  the displacement at the middle of the length $L_{\mathrm N}$ and the equation of motion as:
\begin{equation}\
  v_{\rm{max}}=\frac{\pi(1-\nu)}{16\,b} L_{\mathrm N}^2\,\epsilon\,\omega
\end{equation}
where $\epsilon$ is  the applied strain. This is how we calculated the maximum dislocation speed. Note that in the second regime where the pinning length is $L_{\mathrm i} < L_{\mathrm N}$, the above calculation is no longer strictly valid but the middle of the transition corresponds to $L_{\mathrm i}^2 = L_{\mathrm N}^2/2$ so that our estimate of the dislocation speed is still a good approximation of the exact value.

\begin{figure}
\includegraphics[width=\linewidth]{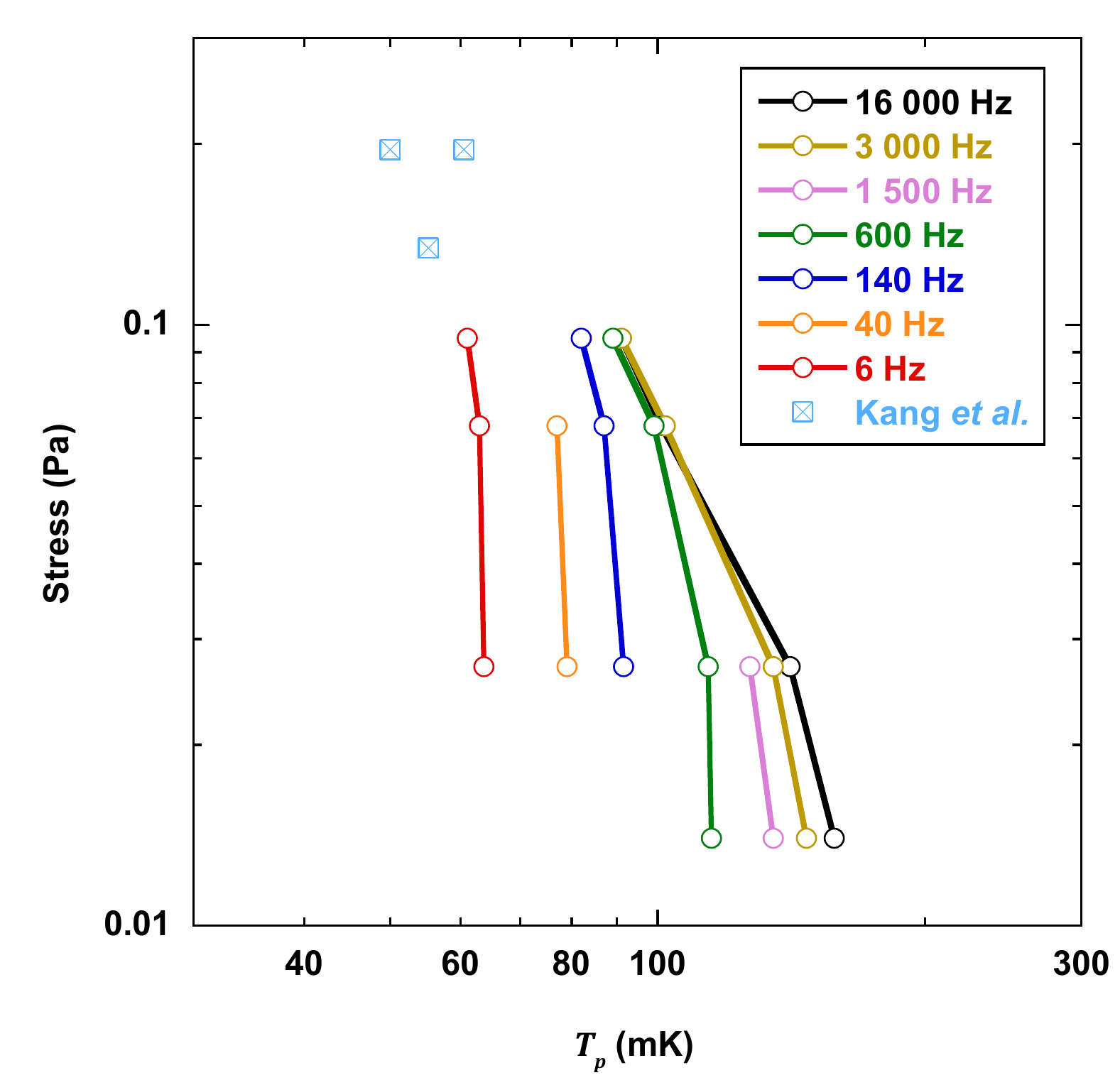}
\caption{(Color online) Here the transition temperature  $T_{\mathrm p}$ is plotted as a function of the stress $\sigma = \mu\epsilon$ resulting from the applied strain $\epsilon$. The stress is calculated from the shear modulus $\mu$ at the transition temperature $T_{\mathrm p}$ for each applied strain, which is roughly the middle of the transition from stiff to soft. The three data points above 0.1~Pa are taken from Kang \etal\cite{Kang13} in the region which they consider as a transition from  fully pinned  to partially pinned dislocations. }
\label{fig:3diagram}
\end{figure}

\begin{center}
\textbf{A tentative interpretation}
\end{center}

As explained above, we have observed a critical speed $v_{\mathrm c}$ where a crossover takes place between a first regime at low frequency and low drive amplitude, which is frequency dependent, and a second regime at higher drive, which is frequency independent. In eq.~\ref{eq:1/Q}, several quantities depend on the concentration of $^3$He atoms bound to the dislocation line, consequently on temperature: the softening $\Delta\mu$, the damping $B$ of the dislocation motion if $^3$He atoms move attached to them, and the pinning length $L$ if it is reduced by the presence of $^3$He atoms on the dislocation. We propose the following interpretation.

\subsection{Critical stress for binding/unbinding to/from \textit{immobile} impurities}

Impurities are bound to edge dislocations with a potential due to the stress field around the dislocation and the volume mismatch between an impurity and the atoms of the crystal.  If a shear stress $\sigma$ is applied to the crystal, the dislocation bows out between pinning points and exerts a force on the impurity which is proportional to the applied stress, to the Burger's vector $b$, and to a pinning length $L$ that is the distance between pinning points on either side of the impurity:

\beq
F= \frac{4b}{\pi} \sigma L
\eeq

If this force exceeds a critical value $F_{c0}$, consequently if the stress exceeds a critical value
$\sigma_{c0}$ (which is inversely proportional to  $L$), the dislocation will detach from the impurity.  This stress-induced breakaway is a purely mechanical process at $T$ = 0 but can be assisted by thermal fluctuations at finite temperature, as explained below. If there are many impurities bound to the dislocation (e.g. at low temperature) then $L$ is essentially the impurity length $L_{\mathrm i}$.  This is small so the breakaway stress is large.  If there are no other bound impurities then $L$ is the longer network pinning length $L_{\mathrm N}$ so the breakaway stress is smaller.  If the crystal is cooled under large stress (starting from high temperature where there are no impurities on the dislocations) then impurities cannot bind and the crystal remains soft to the lowest temperatures.  When the stress is then reduced at low temperature, the first impurity binds to a dislocation at a critical stress corresponding to $L=L_{\mathrm N}$.  This partially pins the dislocation, reducing the length $L$ and raising the critical stress.  Additional impurities can then bind until the impurity length $L_{\mathrm i}$ reaches its equilibrium value and the dislocation is completely pinned.  If all dislocations had the same network length and orientation, they would become pinned at the same stress and the transition from soft to stiff would occur suddenly.  In a real crystal there is a distribution of network lengths so stiffening occurs over a range of stresses, with short dislocations pinning first (at high stress) and longer ones pinning later (at smaller stress).  Modulus measurements made while reducing stress at low temperature can thus be used to determine the distribution of network lengths \cite{Fefferman13}.  The critical stress for binding (when reducing the stress) is controlled by the network length $L_{\mathrm N}$ while the critical stress for unbinding (when increasing the stress) is controlled by the smaller impurity length $L_{\mathrm i}$.  Since the equilibrium density of bound impurities is large at low temperatures, the impurity pinning length $L_{\mathrm i}$ is much smaller than the network length $L_{\mathrm N}$ and the breakaway stress is much larger, which produces the observed hysteresis between the modulus measured while reducing and while increasing the applied stress.

\subsection{Temperature dependence of the binding stress}

At zero temperature, binding and unbinding are purely mechanical processes, controlled by the impurity-dislocation potential (with a depth of order 0.7 K for $^3$He in solid $^4$He).  The energy barrier for unbinding is reduced as the stress increases, disappearing at stress $\sigma_{c0}$.  At finite temperatures the breakaway can be assisted by thermal fluctuations when the energy barrier is small enough.  This gives a critical stress, $\sigma_c(T)$, which decreases at high temperatures.  If measurements are made by cooling a crystal at constant stress (as in our experiments), this leads to a binding temperature $T_{\rm c}(\sigma)$ which decreases as the stress increases, reaching zero at the mechanical breakaway stress $\sigma_{c0}$.  Above this stress the crystal remains soft at all temperatures.  This picture of ``thermally assisted unpinning'' has been described by Teutonico \etal \cite{Teutonico64} and has been used by Kang \etal\cite{Kang13} to explain the amplitude dependence of the shear modulus in solid $^4$He.

\subsection{Critical stress for $^3$He binding in our measurements}

In our measurements shown above, the frequency is held constant while the temperature is reduced at constant strain, i.e. at nearly  constant stress (e.g. the horizontal dashed lines (a) and (b) on Fig. 2, moving from left to right).  For sufficiently high strain, no $^3$He could bind to dislocations and the crystal would remain soft at all temperatures.  At very small strains, we would always be below the critical binding stress corresponding to $L=L_{\mathrm N}$ so $^3$He impurities would attach to dislocations at their equilibrium concentration (which increases exponentially at low temperature:

\beq
X_{\rm 3}^{\rm d} = X_3 \exp{(E/k_{\rm B}T)}
\eeq

At intermediate strains, $^3$He impurities can bind to dislocations only below the stress dependent critical temperature $T_c(\sigma)$ so the crystal stiffens at a temperature which depends on stress but not on frequency.  In Fig. 2, the critical temperatures $T_{\rm c}$ correspond to the dashed vertical lines through the high frequency points.  For example, for the largest strain of 9.5 10$^{-9}$, $T_{\rm c}$  is about 91 mK.  When the strain is reduced, the binding temperature increases, e.g. to about 160 mK at the smallest strain 1.4 10$^{-9}$.  At intermediate temperatures, e.g. 120 mK, the crystal is stiff (at high frequencies) at low stress but soft at high stress.

\subsection{Distinction between ÒbindingÓ and ÒpinningÓ for \textit{mobile} impurities}

In a conventional solid, impurities are immobile at low temperatures so bound impurities can pin the dislocation very effectively.  This restores the solid's intrinsic shear modulus (stiffens the crystal) when the impurity pinning length $L_{\mathrm i}$ becomes shorter than the network length.  For our crystal (with $L_{\mathrm N}$ = 73 $\mu$m, a bulk $^3$He concentration $X_{\rm 3}$ = 300~ppb) this would occur around 240 mK (assuming that the activation energy of 0.67 K is the $^3$He binding energy).  The stiffening temperature would be essentially independent of frequency but would shift to lower temperatures for large stresses.  This explanation breaks down at low frequencies since the measurements in Fig. 2 show that the stiffening transition shifts to lower temperatures at the lowest frequencies.  This is due to the mobility of $^3$He impurities in solid $^4$He.  If impurities can move, then there is a distinction between the binding of an impurity to a dislocation and pinning of the dislocation.  Mobile impurities attached to a dislocation do not necessarily pin it, in contrast to fixed impurities for which binding and pinning are equivalent.  This effect is known in classical crystals, where impurities can move at high temperature via thermally activated diffusion, allowing them to be pulled along with a moving dislocation. Rather than pinning the dislocation, this produces an impurity drag on a moving dislocation.

\subsection{Soft/stiff crossover due to damping from dragged impurities}

At low temperature where phonon scattering disappears and $^3$He binding starts, the damping coefficient B in eq.~2is due to impurities moving with a dislocation.  If each bound $^3$He impurity moves independently, then the damping will be proportional to their density $X_{\rm 3}^{\rm d}$. As a consequence, the relaxation time for dislocation motion ($\tau = BL^2/\pi^2C$) increases exponentially at low temperatures.  If $\tau$ is much longer than the strain oscillation period, the dislocations are effectively immobilized by the damping and the crystal is stiff.  This damping-driven crossover between the soft and stiff states occurs at $\omega\tau = 1$, i.e. at a value of $\tau$ which is inversely proportional to frequency.  A semi log plot of the  frequency  vs. the inverse temperature of the transition $1/T_{\rm p}$ gives a straight line with slope equal to the $^3$He binding energy $E$.  This behavior is seen for the low stress/low frequency data in Fig.~2, giving a binding energy  $E$ = 0.67~K.

\subsection{A critical velocity for dragging $^3$He}

With this picture of $^3$He binding and pinning, we can understand the behavior shown in Fig. 2.  For example, consider the data for strain 6.8~10$^{-9}$.  The binding temperature $T_c$ at this strain (corresponding to the vertical dashed line through the high frequency points) is about 100 mK.  Above this temperature, no $^3$He atoms can bind to the dislocations and the crystal is soft at all frequencies (with a modulus determined by the network length $L_{\mathrm N}$).  Below 100 mK, $^3$He atoms bind to the dislocations at their equilibrium density $L_{\mathrm i}$ ($\ll L_{\mathrm N}$).  At high frequencies (path (b)) these bound impurities immediately stiffen the crystal but at low frequencies (e.g. path (a)) the crystal remains soft.  It is only at a lower temperature (where the drag due to the bound $^3$He impurities immobilizes the dislocations) that we cross over to the stiff state.  This implies that the $^3$He atoms can follow the moving dislocations at low speeds (i.e. at low frequencies) but not at high speeds (high frequencies), the main result of our paper.  Fig.~4, which plots the transition temperatures vs. dislocation speed (rather than vs. frequency) shows that it is this speed which determines whether the $^3$He atoms are able to move with the dislocations.  For dislocation speeds above about 45 $\mu$m/s we see a frequency-dependent, stress-independent transition due to damping from dragged $^3$He impurities.  Above this critical speed, the $^3$He atoms cannot keep up with dislocation motion and so act as nearly fixed pinning centers.  This stiffens the crystal below the binding temperature which is stress-dependent but frequency-independent.

\subsection{A phase diagram}

Figure~5 shows the transitions on a stress vs temperature diagram, in which each point separates a soft from a stiff state at a given frequency.  This diagram allows us to compare our results to those of Kang \etal \cite{Kang13}.  We have calculated the stress by multiplying our applied strain by the value of the shear modulus at the peak dissipation temperatures, i.e. at the temperature of the data points. At high stresses and speeds, the data collapse onto a frequency-independent line along which transition temperatures decrease with increasing stress.  This corresponds to the boundary between the regions Kang \etal describe as "fully pinned" and "partially pinned" on their stress-temperature diagram.  Our data are consistent with their results, although the location of this boundary will depend on crystal quality, i.e. on $L_{\mathrm N}$.  We interpret this  behavior in the same way as Kang \etal , in terms of a critical stress for thermally assisted impurity unbinding.    However, our maximum stress (0.1~Pa) is smaller than their lowest stress (0.2~Pa) and our measurements extend to much lower frequencies (their data was taken at 1~kHz).  This is what allowed us to extend the measurements to a low speed regime and discover a critical speed for $^3$He atoms moving with dislocations.  Below this speed we observe a frequency-dependent stiffening temperature, in agreement with previous work by Syschenko \etal\cite{Syshchenko10} , thus clarifying the apparent discrepancy between the results of the two groups.

\subsection{Migration - inelastic tunneling}

The above physical picture would need a quantitative calculation to be made and compared to experiments, especially for the critical speed and the dissipation amplitude. This is beyond the scope of this article but we have three additional remarks. In 2013, Iwasa~\cite{Iwasa13}  proposed that $^3$He atoms could migrate along the dislocation line. Rojas \etal \cite{Rojas12} actually found some evidence for such a migration that was temperature dependent. When $^3$He atoms stay attached to oscillating dislocations, some migration could occur from the center towards the network nodes that are fixed points. In this scenario, there is a free length at the center of the dislocation, which depends on temperature, amplitude and frequency. One should probably also consider this migration for a future model of the frequency dependence of the dissipation we measured.

In bulk $^4$He crystals of sufficiently high purity and at low temperature, $^3$He atoms are ballistic quasiparticles moving by  coherent tunneling exchange  with $^4$He atoms \cite{Allen82,Sullivan95,Sullivan13}.  There is some uncertainty about the exact value of the exchange frequency $J_{34}/2\pi$. According to the 1982 article by Allen \etal \cite{Allen82} $J_{34}$ should be between $J_{33}$ and $0.1\:J_{33}$ where $J_{33}$ is the exchange frequency in pure $^3$He crystals at the same density. It means $J_{34}/2\pi$ between 0.06 and 0.6~MHz. In 1995, Sullivan  \cite{Sullivan95} proposed $J_{34} /2\pi = 0.42 J_{33}/2\pi$ = 0.23~MHz. The latest estimate by the group of Sullivan \cite{Sullivan13} is $J_{34}/2\pi$~=~1.2~MHz, a  higher value. The average speed of $^3$He quasiparticles is
\beq
<v^2>^{1/2} = 3\sqrt{2} a J_{34}
\label{eq:v3}
\eeq
where $a$ = 0.37~nm is the lattice spacing. To obtain this expression, which is equivalent to the one derived by Sullivan in 1995 \cite{Sullivan95}, we assumed a bandwidth $zh(J_{34}/2\pi)$ with a coordination $z=12$. Given the above estimates of $J_{34}$, we find an average speed between 600~$\mu$m/s and 1.2~cm/s. The critical speed 45~$\mu$m/s could be due to incoherent or inelastic tunneling of $^3$He atoms in the potential well of the dislocation. .

Note that during the capture of $^3$He atoms by dislocations, transverse vibrations are likely to be also emitted in order to release the binding energy. This phenomenon appears similar to the case of  capture by vortices in superfluid $^4$He where Kelvin waves are emitted \cite{Barenghi07,Berloff00,Roberts03}.  Corboz \etal \cite{Corboz08} did not allow for the possible emission of transverse vibrations, which may explain why they came to the conclusion that the capture probability was very small. However, we have observed that, in reality, the capture probability must be large: if we unpin dislocations from $^3$He atoms by applying a large amplitude strain at low temperature, and the large amplitude drive is suddenly released, most of the relaxation of the shear modulus to the value corresponding to pinned dislocations takes a short time of order a few seconds. Of course, a quantitative calculation of the dissipation associated with the motion of a dislocation dressed with moving $^3$He atoms is highly desirable.

\begin{center}
\textbf{Conclusion}
\end{center}

We have found evidence for a  critical dislocation speed $v_{\mathrm c}$. We propose that below $v_{\mathrm c}$ the $^3$He impurities move at the same speed as  the dislocations and damp their motion proportionally to their density on the line. It is also possible that they migrate along  oscillating dislocations if this oscillation is slow enough. Real pinning by $^3$He atoms would occur only above $v_{\mathrm c}$.

This work was supported by grant ERC-AdG 247258 SUPERSOLID,  supported in part by the United States National Science Foundation under Grant No. DMR 0965728,  and by a grant from NSERC Canada.

\bibliography{critical-speed}

\begin{thebibliography}{31}
\expandafter\ifx\csname natexlab\endcsname\relax\def\natexlab#1{#1}\fi
\expandafter\ifx\csname bibnamefont\endcsname\relax
  \def\bibnamefont#1{#1}\fi
\expandafter\ifx\csname bibfnamefont\endcsname\relax
  \def\bibfnamefont#1{#1}\fi
\expandafter\ifx\csname citenamefont\endcsname\relax
  \def\citenamefont#1{#1}\fi
\expandafter\ifx\csname url\endcsname\relax
  \def\url#1{\texttt{#1}}\fi
\expandafter\ifx\csname urlprefix\endcsname\relax\def\urlprefix{URL }\fi
\providecommand{\bibinfo}[2]{#2}
\providecommand{\eprint}[2][]{\url{#2}}

\bibitem[{\citenamefont{Paalanen et~al.}(1981)\citenamefont{Paalanen, Bishop,
  and Dail}}]{Paalanen81}
\bibinfo{author}{\bibfnamefont{M.~A.} \bibnamefont{Paalanen}},
  \bibinfo{author}{\bibfnamefont{D.~J.} \bibnamefont{Bishop}},
  \bibnamefont{and} \bibinfo{author}{\bibfnamefont{H.~W.} \bibnamefont{Dail}},
  \bibinfo{journal}{Phys. Rev. Lett.} \textbf{\bibinfo{volume}{46}},
  \bibinfo{pages}{664} (\bibinfo{year}{1981}).

\bibitem[{\citenamefont{Day and Beamish}(2007)}]{Day07}
\bibinfo{author}{\bibfnamefont{J.}~\bibnamefont{Day}} \bibnamefont{and}
  \bibinfo{author}{\bibfnamefont{J.}~\bibnamefont{Beamish}},
  \bibinfo{journal}{Nature} \textbf{\bibinfo{volume}{450}},
  \bibinfo{pages}{853} (\bibinfo{year}{2007}).

\bibitem[{\citenamefont{Day et~al.}(2010)\citenamefont{Day, Syshchenko, and
  Beamish}}]{Day10}
\bibinfo{author}{\bibfnamefont{J.}~\bibnamefont{Day}},
  \bibinfo{author}{\bibfnamefont{O.}~\bibnamefont{Syshchenko}},
  \bibnamefont{and} \bibinfo{author}{\bibfnamefont{J.}~\bibnamefont{Beamish}},
  \bibinfo{journal}{Phys. Rev. Lett.} \textbf{\bibinfo{volume}{104}},
  \bibinfo{pages}{075302} (\bibinfo{year}{2010}).

\bibitem[{\citenamefont{Syshchenko et~al.}(2010)\citenamefont{Syshchenko, Day,
  and Beamish}}]{Syshchenko10}
\bibinfo{author}{\bibfnamefont{O.}~\bibnamefont{Syshchenko}},
  \bibinfo{author}{\bibfnamefont{J.}~\bibnamefont{Day}}, \bibnamefont{and}
  \bibinfo{author}{\bibfnamefont{J.}~\bibnamefont{Beamish}},
  \bibinfo{journal}{Phys. Rev. Lett.} \textbf{\bibinfo{volume}{104}},
  \bibinfo{pages}{195301} (\bibinfo{year}{2010}).

\bibitem[{\citenamefont{Rojas et~al.}(2010)\citenamefont{Rojas, Haziot, Babst,
  Balibar, and Maris}}]{Rojas10}
\bibinfo{author}{\bibfnamefont{X.}~\bibnamefont{Rojas}},
  \bibinfo{author}{\bibfnamefont{A.}~\bibnamefont{Haziot}},
  \bibinfo{author}{\bibfnamefont{V.}~\bibnamefont{Babst}},
  \bibinfo{author}{\bibfnamefont{S.}~\bibnamefont{Balibar}}, \bibnamefont{and}
  \bibinfo{author}{\bibfnamefont{H.~J.} \bibnamefont{Maris}},
  \bibinfo{journal}{Phys. Rev. Lett.} \textbf{\bibinfo{volume}{105}},
  \bibinfo{pages}{145302} (\bibinfo{year}{2010}).

\bibitem[{\citenamefont{Varoquaux}(2012)}]{Varoquaux12}
\bibinfo{author}{\bibfnamefont{E.}~\bibnamefont{Varoquaux}},
  \bibinfo{journal}{Phys. Rev. B} \textbf{\bibinfo{volume}{86}},
  \bibinfo{pages}{064524} (\bibinfo{year}{2012}).

\bibitem[{\citenamefont{Haziot et~al.}(2013{\natexlab{a}})\citenamefont{Haziot,
  Rojas, Fefferman, Beamish, and Balibar}}]{Haziot13a}
\bibinfo{author}{\bibfnamefont{A.}~\bibnamefont{Haziot}},
  \bibinfo{author}{\bibfnamefont{X.}~\bibnamefont{Rojas}},
  \bibinfo{author}{\bibfnamefont{A.}~\bibnamefont{Fefferman}},
  \bibinfo{author}{\bibfnamefont{J.}~\bibnamefont{Beamish}}, \bibnamefont{and}
  \bibinfo{author}{\bibfnamefont{S.}~\bibnamefont{Balibar}},
  \bibinfo{journal}{Phys. Rev. Lett.} \textbf{\bibinfo{volume}{110}},
  \bibinfo{pages}{035301} (\bibinfo{year}{2013}{\natexlab{a}}).

\bibitem[{\citenamefont{Haziot et~al.}(2013{\natexlab{b}})\citenamefont{Haziot,
  Fefferman, Beamish, and Balibar}}]{Haziot13b}
\bibinfo{author}{\bibfnamefont{A.}~\bibnamefont{Haziot}},
  \bibinfo{author}{\bibfnamefont{A.}~\bibnamefont{Fefferman}},
  \bibinfo{author}{\bibfnamefont{J.}~\bibnamefont{Beamish}}, \bibnamefont{and}
  \bibinfo{author}{\bibfnamefont{S.}~\bibnamefont{Balibar}},
  \bibinfo{journal}{Phys. Rev. B} \textbf{\bibinfo{volume}{89}},
  \bibinfo{pages}{060509(R)} (\bibinfo{year}{2013}{\natexlab{b}}).

\bibitem[{\citenamefont{Beamish et~al.}(2012)\citenamefont{Beamish, Haziot,
  Rojas, Fefferman, and Balibar}}]{Beamish12}
\bibinfo{author}{\bibfnamefont{J.}~\bibnamefont{Beamish}},
  \bibinfo{author}{\bibfnamefont{A.}~\bibnamefont{Haziot}},
  \bibinfo{author}{\bibfnamefont{X.}~\bibnamefont{Rojas}},
  \bibinfo{author}{\bibfnamefont{A.}~\bibnamefont{Fefferman}},
  \bibnamefont{and} \bibinfo{author}{\bibfnamefont{S.}~\bibnamefont{Balibar}},
  \bibinfo{journal}{Phys. Rev. B} \textbf{\bibinfo{volume}{85}},
  \bibinfo{pages}{180501} (\bibinfo{year}{2012}).

\bibitem[{\citenamefont{Maris}(2012)}]{Maris12}
\bibinfo{author}{\bibfnamefont{H.}~\bibnamefont{Maris}},
  \bibinfo{journal}{Phys. Rev. B} \textbf{\bibinfo{volume}{86}},
  \bibinfo{pages}{020502} (\bibinfo{year}{2012}).

\bibitem[{\citenamefont{Reppy et~al.}(2012)\citenamefont{Reppy, Mi, Justin, and
  Mueller}}]{Reppy12-175}
\bibinfo{author}{\bibfnamefont{J.}~\bibnamefont{Reppy}},
  \bibinfo{author}{\bibfnamefont{X.}~\bibnamefont{Mi}},
  \bibinfo{author}{\bibfnamefont{A.}~\bibnamefont{Justin}}, \bibnamefont{and}
  \bibinfo{author}{\bibfnamefont{E.}~\bibnamefont{Mueller}},
  \bibinfo{journal}{J. Low Temp. Phys.} \textbf{\bibinfo{volume}{168}},
  \bibinfo{pages}{175} (\bibinfo{year}{2012}).

\bibitem[{\citenamefont{Chan}(2012)}]{Chan12}
\bibinfo{author}{\bibfnamefont{M.}~\bibnamefont{Chan}}, \bibinfo{journal}{Phys.
  Rev. B}  (\bibinfo{year}{2012}).

\bibitem[{\citenamefont{Kim and Chan}(2004{\natexlab{a}})}]{Kim04a}
\bibinfo{author}{\bibfnamefont{E.}~\bibnamefont{Kim}} \bibnamefont{and}
  \bibinfo{author}{\bibfnamefont{M.~H.~W.} \bibnamefont{Chan}},
  \bibinfo{journal}{Nature} \textbf{\bibinfo{volume}{427}},
  \bibinfo{pages}{225} (\bibinfo{year}{2004}{\natexlab{a}}).

\bibitem[{\citenamefont{Kim and Chan}(2004{\natexlab{b}})}]{Kim04b}
\bibinfo{author}{\bibfnamefont{E.}~\bibnamefont{Kim}} \bibnamefont{and}
  \bibinfo{author}{\bibfnamefont{M.~H.~W.} \bibnamefont{Chan}},
  \bibinfo{journal}{Science} \textbf{\bibinfo{volume}{305}},
  \bibinfo{pages}{1941} (\bibinfo{year}{2004}{\natexlab{b}}).

\bibitem[{\citenamefont{Iwasa}(2010)}]{Iwasa10}
\bibinfo{author}{\bibfnamefont{I.}~\bibnamefont{Iwasa}},
  \bibinfo{journal}{Phys. Rev. B} \textbf{\bibinfo{volume}{81}},
  \bibinfo{pages}{104527} (\bibinfo{year}{2010}).

\bibitem[{\citenamefont{Kang et~al.}(2013)\citenamefont{Kang, Kim, Kim, and
  Kim}}]{Kang13}
\bibinfo{author}{\bibfnamefont{E.}~\bibnamefont{Kang}},
  \bibinfo{author}{\bibfnamefont{D.}~\bibnamefont{Kim}},
  \bibinfo{author}{\bibfnamefont{H.}~\bibnamefont{Kim}}, \bibnamefont{and}
  \bibinfo{author}{\bibfnamefont{E.}~\bibnamefont{Kim}},
  \bibinfo{journal}{Phys. Rev. B} \textbf{\bibinfo{volume}{87}},
  \bibinfo{pages}{094512} (\bibinfo{year}{2013}).

\bibitem[{\citenamefont{Granato and Lucke}(1956)}]{Granato56}
\bibinfo{author}{\bibfnamefont{A.}~\bibnamefont{Granato}} \bibnamefont{and}
  \bibinfo{author}{\bibfnamefont{K.}~\bibnamefont{Lucke}}, \bibinfo{journal}{J.
  Appl. Phys.} \textbf{\bibinfo{volume}{27}}, \bibinfo{pages}{583}
  (\bibinfo{year}{1956}).

\bibitem[{\citenamefont{Crepeau et~al.}(1971)\citenamefont{Crepeau, Heybey,
  Lee, and S.A.}}]{Crepeau}
\bibinfo{author}{\bibfnamefont{R.}~\bibnamefont{Crepeau}},
  \bibinfo{author}{\bibfnamefont{O.}~\bibnamefont{Heybey}},
  \bibinfo{author}{\bibfnamefont{D.}~\bibnamefont{Lee}}, \bibnamefont{and}
  \bibinfo{author}{\bibfnamefont{S.}~\bibnamefont{S.A.}},
  \bibinfo{journal}{Phys. Rev. A} \textbf{\bibinfo{volume}{3}},
  \bibinfo{pages}{1162} (\bibinfo{year}{1971}).

\bibitem[{\citenamefont{Greywall}(1977)}]{Greywall}
\bibinfo{author}{\bibfnamefont{D.}~\bibnamefont{Greywall}},
  \bibinfo{journal}{Phys. Rev. B} \textbf{\bibinfo{volume}{16}},
  \bibinfo{pages}{5127} (\bibinfo{year}{1977}).

\bibitem[{\citenamefont{Friedel}(1964)}]{Friedel64}
\bibinfo{author}{\bibfnamefont{J.}~\bibnamefont{Friedel}},
  \emph{\bibinfo{title}{{Dislocations}}} (\bibinfo{publisher}{Pergamon},
  \bibinfo{address}{New York}, \bibinfo{year}{1964}).

\bibitem[{\citenamefont{Fefferman et~al.}(2013)\citenamefont{Fefferman, Souris,
  Haziot, Beamish, and Balibar}}]{Fefferman13}
\bibinfo{author}{\bibfnamefont{A.}~\bibnamefont{Fefferman}},
  \bibinfo{author}{\bibfnamefont{F.}~\bibnamefont{Souris}},
  \bibinfo{author}{\bibfnamefont{A.}~\bibnamefont{Haziot}},
  \bibinfo{author}{\bibfnamefont{J.}~\bibnamefont{Beamish}}, \bibnamefont{and}
  \bibinfo{author}{\bibfnamefont{S.}~\bibnamefont{Balibar}},
  \bibinfo{journal}{to appear}  (\bibinfo{year}{2013}).

\bibitem[{\citenamefont{Teutonico et~al.}(1964)\citenamefont{Teutonico,
  Granato, and Lucke}}]{Teutonico64}
\bibinfo{author}{\bibfnamefont{L.}~\bibnamefont{Teutonico}},
  \bibinfo{author}{\bibfnamefont{A.}~\bibnamefont{Granato}}, \bibnamefont{and}
  \bibinfo{author}{\bibfnamefont{K.}~\bibnamefont{Lucke}}, \bibinfo{journal}{J
  Appl Phys} \textbf{\bibinfo{volume}{35}}, \bibinfo{pages}{2732}
  (\bibinfo{year}{1964}).

\bibitem[{\citenamefont{Iwasa}(2013)}]{Iwasa13}
\bibinfo{author}{\bibfnamefont{I.}~\bibnamefont{Iwasa}}, \bibinfo{journal}{J.
  Low Temp. Phys.} \textbf{\bibinfo{volume}{171}}, \bibinfo{pages}{30}
  (\bibinfo{year}{2013}).

\bibitem[{\citenamefont{Rojas et~al.}(2012)\citenamefont{Rojas, Haziot, and
  Balibar}}]{Rojas12}
\bibinfo{author}{\bibfnamefont{X.}~\bibnamefont{Rojas}},
  \bibinfo{author}{\bibfnamefont{A.}~\bibnamefont{Haziot}}, \bibnamefont{and}
  \bibinfo{author}{\bibfnamefont{S.}~\bibnamefont{Balibar}},
  \bibinfo{journal}{J. Physics: Conf .Series} \textbf{\bibinfo{volume}{400}},
  \bibinfo{pages}{012062} (\bibinfo{year}{2012}).

\bibitem[{\citenamefont{Allen et~al.}(1982)\citenamefont{Allen, Richards, and
  Schratter}}]{Allen82}
\bibinfo{author}{\bibfnamefont{A.}~\bibnamefont{Allen}},
  \bibinfo{author}{\bibfnamefont{M.}~\bibnamefont{Richards}}, \bibnamefont{and}
  \bibinfo{author}{\bibfnamefont{J.}~\bibnamefont{Schratter}},
  \bibinfo{journal}{J. Low Temp. Phys.} \textbf{\bibinfo{volume}{47}},
  \bibinfo{pages}{289} (\bibinfo{year}{1982}).

\bibitem[{\citenamefont{Sullivan}(1995)}]{Sullivan95}
\bibinfo{author}{\bibfnamefont{N.}~\bibnamefont{Sullivan}},
  \bibinfo{journal}{Appl. Magn. Reson.} \textbf{\bibinfo{volume}{8}},
  \bibinfo{pages}{361} (\bibinfo{year}{1995}).

\bibitem[{\citenamefont{Kim et~al.}(2013)\citenamefont{Kim, Huan, Yin, Xia,
  Candela, and Sullivan}}]{Sullivan13}
\bibinfo{author}{\bibfnamefont{S.~S.} \bibnamefont{Kim}},
  \bibinfo{author}{\bibfnamefont{C.}~\bibnamefont{Huan}},
  \bibinfo{author}{\bibfnamefont{L.}~\bibnamefont{Yin}},
  \bibinfo{author}{\bibfnamefont{J.~S.} \bibnamefont{Xia}},
  \bibinfo{author}{\bibfnamefont{D.}~\bibnamefont{Candela}}, \bibnamefont{and}
  \bibinfo{author}{\bibfnamefont{N.}~\bibnamefont{Sullivan}},
  \bibinfo{journal}{arXiv:1301.3929}
  \textbf{\bibinfo{volume}{cond-mat.mtrl-sci}} (\bibinfo{year}{2013}).

\bibitem[{\citenamefont{Barenghi et~al.}(2007)\citenamefont{Barenghi,
  Kivotides, and Sergeev}}]{Barenghi07}
\bibinfo{author}{\bibfnamefont{C.}~\bibnamefont{Barenghi}},
  \bibinfo{author}{\bibfnamefont{D.}~\bibnamefont{Kivotides}},
  \bibnamefont{and} \bibinfo{author}{\bibfnamefont{Y.}~\bibnamefont{Sergeev}},
  \bibinfo{journal}{J. Low Temp. Phys} \textbf{\bibinfo{volume}{148}},
  \bibinfo{pages}{293} (\bibinfo{year}{2007}).

\bibitem[{\citenamefont{Berloff and Roberts}(2000)}]{Berloff00}
\bibinfo{author}{\bibfnamefont{N.}~\bibnamefont{Berloff}} \bibnamefont{and}
  \bibinfo{author}{\bibfnamefont{P.}~\bibnamefont{Roberts}},
  \bibinfo{journal}{Phys. Rev. B} \textbf{\bibinfo{volume}{63}},
  \bibinfo{pages}{024510} (\bibinfo{year}{2000}).

\bibitem[{\citenamefont{Roberts}(2003)}]{Roberts03}
\bibinfo{author}{\bibfnamefont{P.}~\bibnamefont{Roberts}},
  \bibinfo{journal}{Proc. R. Soc. Lond. A} \textbf{\bibinfo{volume}{459}},
  \bibinfo{pages}{331} (\bibinfo{year}{2003}).

\bibitem[{\citenamefont{Corboz et~al.}(2008)\citenamefont{Corboz, Pollet,
  ProkofÕev, and Troyer}}]{Corboz08}
\bibinfo{author}{\bibfnamefont{P.}~\bibnamefont{Corboz}},
  \bibinfo{author}{\bibfnamefont{L.}~\bibnamefont{Pollet}},
  \bibinfo{author}{\bibfnamefont{N.}~\bibnamefont{ProkofÕev}},
  \bibnamefont{and} \bibinfo{author}{\bibfnamefont{M.}~\bibnamefont{Troyer}},
  \bibinfo{journal}{Phys. Rev. Lett.} \textbf{\bibinfo{volume}{101}},
  \bibinfo{pages}{155302} (\bibinfo{year}{2008}).

\end{thebibliography}

\end{document}